# Electrochemical Strain Microscopy of Silica Glasses

R. Proksch[1]

Asylum Research, an Oxford Instruments Company, 6310 Hollister Avenue, Santa Barbara, CA 93117 USA

Piezoresponse Force Microscopy (PFM) and Electrochemical Strain Microscopy (ESM) are two related techniques that have had considerable success in nano-scale probing of functional material properties. Both measure the strain of the sample in response to a localized electric field beneath a sharp conductive tip. In this work, a collection of commercially available glass samples were measured with a variety of Si cantilevers coated with different conductive metals. In some cases, these glasses showed significant hysteresis loops, similar in appearance to those measured on ferroelectric materials with spontaneous permanent electric dipoles. The magnitude of the electrochemical strain and hysteresis correlated well with the molar percentage of sodium in the glass material, with high sodium (soda-lime) glass showing large hysteresis and fused silica (pure $SiO_2$) showing essentially no hysteresis. The "elephant-ear" shape of the hysteresis loops correlated well with it originating from relaxation behavior – an interpretation verified by observing the temperature dependent relaxation of the ESM response. Cation mobility in a disordered glass should have a low diffusion constant. To evaluate this diffusion constant, the temperature of the glass was varied between room temperature to ~200C. Vanishing hysteresis as the temperature increased was associated with a decrease in the relaxation time of the electrochemical response. The hysteretic behavior changed drastically in this temperature range, consistent with bound surface water playing a large role in the relaxation. This demonstrates the ability of ESM to differentiate cationic concentrations in a range of silica glasses. In addition, since glass is a common sample substrate for, this provides some clear guidance for avoiding unwanted substrate crosstalk effects in piezoresponse and electrochemical strain response measurements.

---

[1] Roger.Proksch@oxinst.com

The high spatial resolution and direct sensitivity to localized strain has enabled the atomic force microscope (AFM)[1] technique to be applied to many of these electromechanical materials. Initially, most of this local probing was of piezo- and ferro-electric materials with PFM.[2, 3, 4, 5, 6] Recently, ESM has been applied to a range of non-piezoelecric but still electromechanically active materials.[7, 8, 9] Beyond this, analytic electrochemical sensors are ubiquitous in a range of industries, from medical diagnostics, process monitoring in chemical plants to environmental monitoring equipment. Unwanted electrochemical phenomena such as corrosion[10] have an enormous economic impact. Finally, biological molecules and systems exhibit an enormous range of electromechanical responses sometime interpreted as piezo- or ferro-electricity. [11, 12, 13]

One of the earliest local probe electromechanical techniques was PFM,[14, 15] a well-established technique that measures the local polarization strength and direction of ferro- and piezo-electric materials. In PFM, the cantilever tip is placed in contact with the surface while the tip-sample voltage is modulated with a periodic bias that leads to deformations in the sample surface, which in turn is measured by the cantilever. When the response of the cantilever is dominated by the piezoelectric contribution, the phase of the electromechanical response of the surface, $\phi$ is dependent on the polarization direction. The observation of localized hysteresis is a hallmark measurement of this technique and is generally considered a "smoking gun" for nanoscale ferroelectricity.[16, 17, 18, 19, 20, 21, 22, 23, 24] Localized hysteresis loops are typically measured by ramping or stepping a DC voltage in addition to a small AC excitation.

In addition to the piezoelectric strains, capacitive coupling between the tip and sample can lead to large responses at the drive frequency.[25, 26, 27, 28] Since this response occurs at the same frequency as the piezoresponse, it is a background signal that must be either overcome or otherwise eliminated for quantitative PFM measurements.[29, 30] In general, eliminating these effects remains a significant challenge. In the case of hysteresis loops, the Switching Spectroscopy PFM (SSPFM) technique developed by Jesse et al.[31] mitigates the effects of electrostatics by making remnant measurements in a zero applied voltage.

Electrochemical Strain Microscopy is a developing technique where localized bias-induced changes in the local strain of a surface, caused by ion transport, usually in a solid electrolyte material is measured with a cantilever.[32, 33] As with PFM, an oscillating potential is applied to the tip and the deflection at the drive frequency is analyzed to determine both the amplitude and phase of the electromechanical response.

Glass is sometimes thought of as an inert and insulating dielectric material. Because of this, it is used as an inexpensive substrate for a variety of thin films, including piezo- and ferro-electric materials such as PZT[34] and even biological materials.[35] In general however, glasses are electrochemically active. Glass can be used as electrodes in batteries, it can corrode and glass electrodes are important functional components one of the most common analytical tool – pH and ion-selective membrane probes.[36, 37] Silicate Oxide glasses are comprised of a disordered network of $SiO_2$ and other cations, typically $Na^+$, $K^+$, $Li^+$, $Ca^{2+}$ or $B^{2+}$, depending on the glass.[38,39] The ionic velocity of these cations in response to an electric field or other gradients through the disordered $SiO_2$ matrix is very low and limited by the disordered nature of the glass structure.[40]

In this work, we have explored the electrochemical response of a variety of silicate glasses (listed in Table 1) interacting with Pt[41] and Iridium[42] coated cantilevers. All measurements were conducted with a MFP-3D AFM (Asylum Research, Santa Barbara, CA) equipped with a HVA-220 high voltage amplifier. Operating on the contact resonance frequency enhances the sensitivity of the measurement while simultaneously providing additional information regarding the sample stiffness and dissipation.

The contact resonance can be used with either dual AC resonance tracking (DART) mode[38] of Band Excitation(BE).[43] This work used DART, but BE is expected to produce similar results. The relative humidity of the lab during the course of the measurements ranged from 35% to 40%. Bias dependent amplitude and phase responses at the contact resonance frequency were measured on the clean glass surfaces. A typical experiment was to place the tip at various points on the surface in contact while the bias is changed to perform a hysteresis loop. At the same time, DART measurements of the contact resonance measure the electrochemical strain response. Before and after these loop(s) are performed, the same cantilever can scan over the region in a gentle tapping mode to examine the area for modifications.[44] Interestingly, the specific location of the contact ground did not seem to affect the results significantly. The results presented here were taken with an electrode on the bottom but grounding the top surface of the glass several mm away from the cantilever contact area gave identical results.

The glass samples were cleaned before imaging and hysteresis loop measurements using the following protocol: vigorous scrubbing with a dish detergent followed by a thorough rinsing in distilled water, 10 minutes of sonication in acetone followed by water rinsing, 10 minutes of sonication in ethanol followed by water rinsing and a final 10 minute sonication in distilled water. After the final sonication step, the glass was rinsed in water and dried with compressed clean nitrogen gas. All of the glass samples were tested and in all cases were observed to be highly hydrophilic – a distilled water droplet spread out evenly over the surface.

A typical series of 30 volt amplitude hysteresis loops made on a cleaned glass surface are shown in Figure 1. Within the limits of the signal to noise, the loop shape was repeatable. Small particles are visible on the surface of Figure 1(b) in proximity of the tip location during the hysteresis loops – clearly implying electrochemical reactions. In addition, the particle size in Figure 1(b) scales with the number of bias cycles. Although particle growth in Figure 1(b) appear somewhat regular, the existence of particles on any given glass surface was highly variable, dependent on the cantilever used. This is consistent with expected variations in the details of the localized electric field emanating from the tip of the cantilever.

Figure 1 (c) shows amplitude butterfly loops and 1(d) the associated phase response as the maximum bias voltage value was increased. Note that the shape of the first amplitude loop in Figure 1(c) is similar to the 30 V loops shown in Figure 1(a). As the maximum voltage was increased, the area enclosed by the loop increased and the loop matured into a full-blown butterfly loop with rounded lobes on the top. With the exception of the rounded "elephant ear" shape, the loops are what we might expect from ferroelectric switching measurements. Figure 1(d) shows a full 180 degree phase shift, also as might be expected from a ferroelectric switching experiment. All of the loops in Figures 1(c) and 1(d) were made at a rate of 5 seconds/loop. Because the loop time was held constant while the maximum bias was increased, the bias ramp rate increased proportional to the maximum amplitude. Unlike the results we might expect from ferroelectric materials, the coercive voltage (where the amplitude goes through a minimum) and the area enclosed by the loop is voltage sweep rate dependent in time range accessible to the AFM.

Cationic diffusion processes such as those leading to electrochemical strain effects are expected to be temperature dependent. To study the role of temperature for ESM in these glass samples, loops were measured as a function of temperature using a polymer heater from room temperature up to 200C. Figure 2 shows ESM amplitude butterfly loops versus bias voltage ranging from -50V to 50V for a variety of glasses as a function of temperature. This range was chosen because it was large enough to fully open the loops in the soda-lime glass of Figure 1(c). The soda-lime loops shown in Figures 2(a-c) showed vanishing hysteresis as the temperature approached 200C. This is in stark contrast to the fused silica shown in Figure 2(d), where the hysteresis was smaller than the experimental noise levels at all temperatures. The observed hysteresis is dependent on the material studied, with higher sodium content glass exhibiting larger hysteresis, consistent with it originating from cationic diffusion in response to the tip electric field. This provides a cautionary lesson for use of glasses for sample supports, at least when electric field is penetrates to the glass layer. For thick samples, the effect of the substrate is expected to be negligible [35]. The hysteresis vanishing above 100C is consistent with the surface water playing a role in the ESM signal generation.[45]

Another interesting observation from Figures 2(a) and 2(b) is that the hysteresis loops measured over the Fisher soda-lime glass depended on the metal coating of the cantilever, with the Pt tip showing a larger response than the Ir tip. While there is significant variability in the hysteresis loops from tip to tip, this trend was verified by repeating the measurements with a series of five Pt tips and five Ir tips. This is consistent with the electro-catalytic activity of Pt being larger than that of Ir. It also suggests that it may be possible to evaluate the electrocatalytic differences between different materials using a reference electrode material.

Figure 3 shows the detailed spectroscopic ESM amplitude signal versus time response that was used to calculate the hysteresis loops of Figures 1 and 2. The on and off portions of the excitation voltage were 100ms as shown in Figure 3(a). The software routine used to calculate hysteresis averages the amplitude for half the on time and half the off time as shown in Figure 3(a), avoiding transient effects from the transition between the two states. Figure 3(b) is from the loops of Figure 2(a), 3(c) from the loops of 2(c) and 3(d) from 2(d) respectively. In particular, a small portion of the spectroscopic ESM amplitude signal used to calculate the loops of Figure 2(a) are shown in Figure 3(b). The variation in relaxation times is apparent from the time data and is also explicitly plotted in Figure 3(e) versus the sample temperature. Note that when the measurement times are much greater than the relaxation times ($t_{off}, t_{on} \geq 10\tau$ where $\tau$ is the ESM relaxation time), the hysteresis vanishes. Note that ferroelectric materials will also exhibit slow dynamics. Teasing those slow relaxation phenomena will be difficult to unambiguously observe and may require additional measurements such as time-lapse domain imaging.

In conclusion, commercially available high sodium and fused silica glass samples were measured with a variety of Si cantilevers coated with different conductive metals. The high sodium glasses showed both significant electrochemical strain amplitudes and hysteresis while the fused silica samples showed very little response and hysteresis was below detectable levels. The hysteresis loops on the sodium glass materials appeared similar to those measured on ferroelectric materials, except for their particular "elephant ear" shape. The "elephant ear" appears in hysteresis loops originating from relaxation processes in the ESM signal. This observation was confirmed by the time dependent spectroscopic data. In addition to demonstrating high spatial resolution nanoscale electrochemical activity mapping capabilities, there are some clear cautionary lessons when using electrochemically active substrates for PFM and ESM. In the materials studied here, the more electrochemically active glasses (with the higher monovalent cation concentrations) showed hysteresis. There should be similar behavior in other materials with mobile ions. Hysteresis measurement times can be adjusted to mitigate the relaxation hysteresis, for example by slowing down the bias sweep rate so that the measurement time is much greater than the electrochemical relaxation times. In this case, the hysteresis is minimized. Finally, for PFM and ESM measurements, these results strongly suggest using care in selecting the substrate. In the case of the samples measured here, a low-cation concentration fused silica of the sort listed in Table 1 would better avoid unwanted crosstalk effects.

**Table 1**

| Glass Type | Na$_2$O | K$_2$O | CaO | Vendor | Measurements |
|---|---|---|---|---|---|
| Fused Silica | 0% | 0% | 0% | Edmund | Figures 2(d) and 3(d) |
| Soda-Lime Float Glass | 14.3% | 1.2% | 6.4% | Fisher Scientific | Figures 1(a), 1(b), 2(a), 2(b) and 3(b) |
| Soda-Lime Float Glass | 13.9% | 0.6% | 8.4% | Edmund | Figures 2(c) and 3(c) |

**Figures**

Figure 1: (a) shows a series of 30-volt amplitude hysteresis loops measured with a stepped ramp bias drive from Jesse et al. measured above on a soda-lime glass sample. The amplitude loops are color-coded to locations on an image (b) of the sample made after the loop measurements. The number of cycles in the loops progresses from 1 (red), 1.5 (orange) 2 (green) 2.5 (blue) 3 (magenta). The shape of the loops had a very high degree of repeatability in all cases. Small particles remained behind in the proximity of the tip. (c) shows amplitude and (d) the associated phase loops for increasing voltage ranges over a different region of the same sample as the data in (a) and (b). The increasingly "elephant-ear" shaped amplitude butterfly loops are shown in (c) as a color-coded function of the maximum ramp voltage. The phase loops (d) show a full 180 degree shift for the larger voltage ranges. The loop measurement time is held constant at 5 seconds/loop in each case. This results in different voltage ramp rates of 48 (red) 64 (orange), 80 (green), 96 (blue), 112 (purple) and 128 V/sec (black).

Figure 2: Temperature color-coded ESM amplitude butterfly loops versus bias voltage ranging from -50V to 50V for (a) Soda-lime (Fisher), Pt tip, (b) Soda-lime (Fisher) Ir tip, (c) Soda-lime (Edmunds) Pt tip and (d) Fused Silica (Edmunds) Pt tip. While the soda lime glasses showed some hysteresis beyond 100C, any hysteresis in the fused silica glass was smaller than the experimental noise levels.

Figure 3: Figure 3(a) shows the stepped bias voltage used to excite the ESM response in the sample. $t_{on}$ and $t_{off}$ show the averaging times for the amplitude displayed in the hysteresis loops of Figures 1 and 2. (b) shows the amplitude response for the high sodium float glass (Fisher) hysteresis loops shown in Figure 2(a). Figure (c) shows the same curves used to generate the hysteresis loops of Figure 2(c), the slightly lower sodium content float glass (Edmunds). In this case, the amplitudes were roughly half the magnitude. In addition, the amplitude increased as a function temperature for both the soda-

lime glass samples. (d) shows a subset of the time data used to generate the Fused Silica sample hysteresis loops of Figure 2(d). In this case, the response is not significantly above the noise level for all temperatures. (e) shows the time constants extracted from some of the relaxation curves plotted as a function of temperature for the two soda-lime glasses. In addition, The 50ms averaging time for the hysteresis loops is shown as a solid horizontal black line. As the relaxation time of the heated glasses drops below this measurement time, the hysteresis vanishes. All the data in this Figure was taken with the same Pt coated cantilever.

**Acknowledgements**

Portions of this work were sparked after a conversation with G. Sheet. Conversations with Sergei Kalinin at ORNL and Jinagyu Li at the University of Washington were invaluable during the preparation of the manuscript.

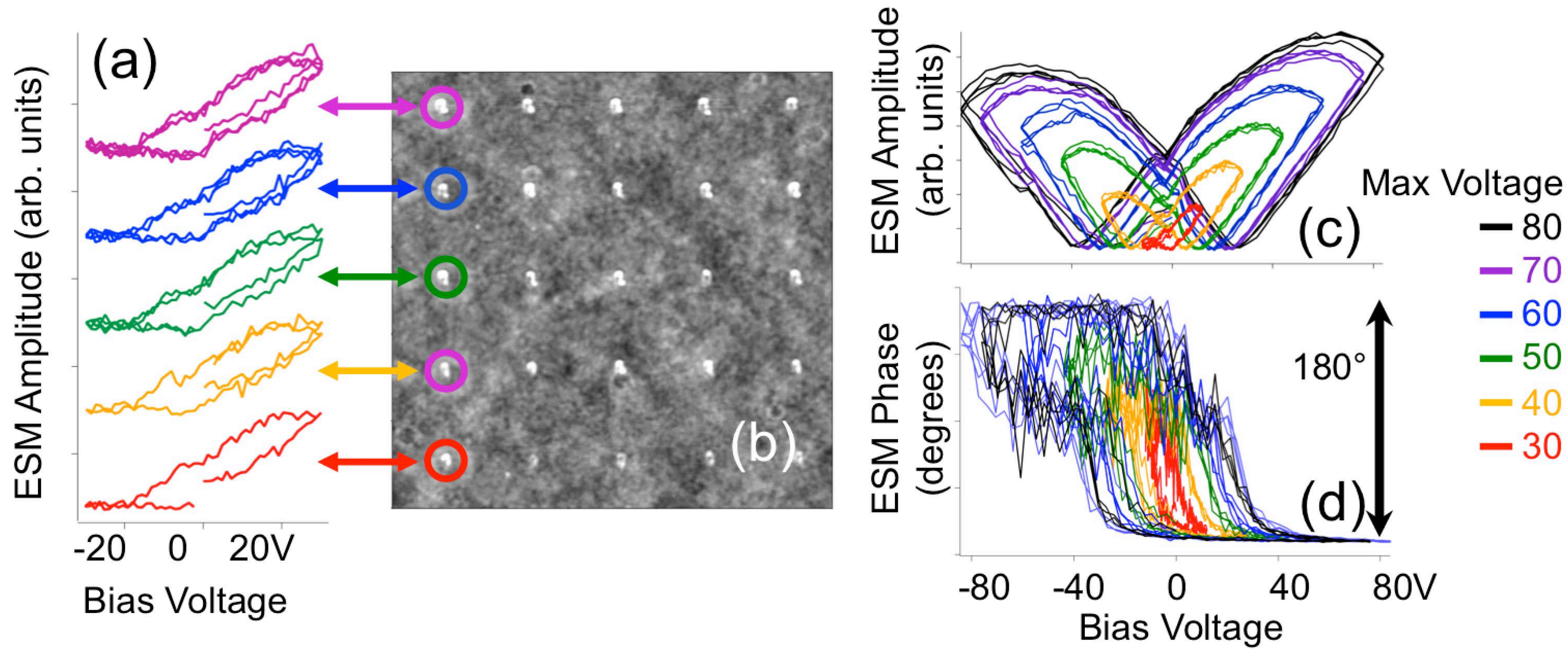

Figure 1, Proksch

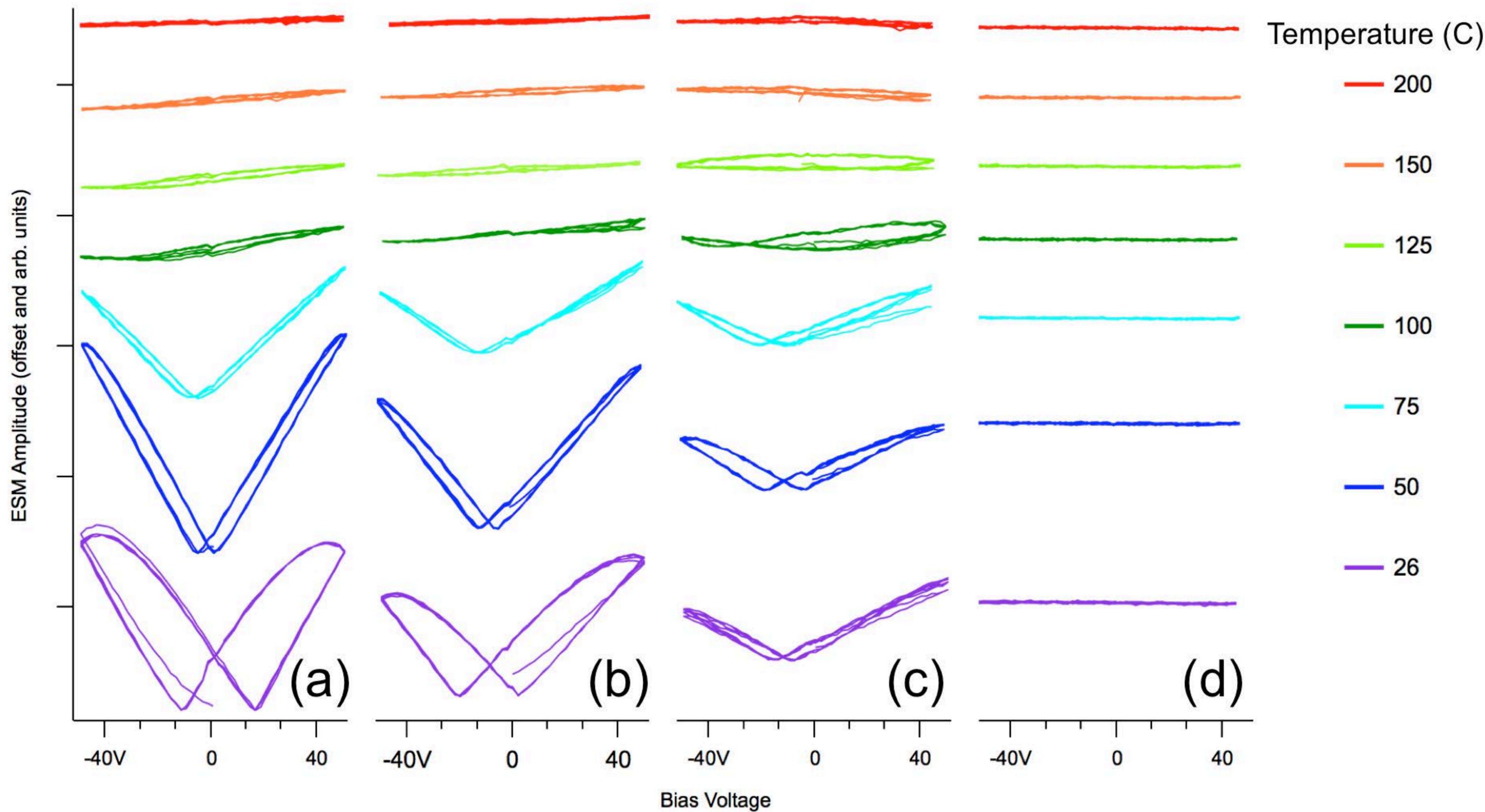

Figure 2, Proksch

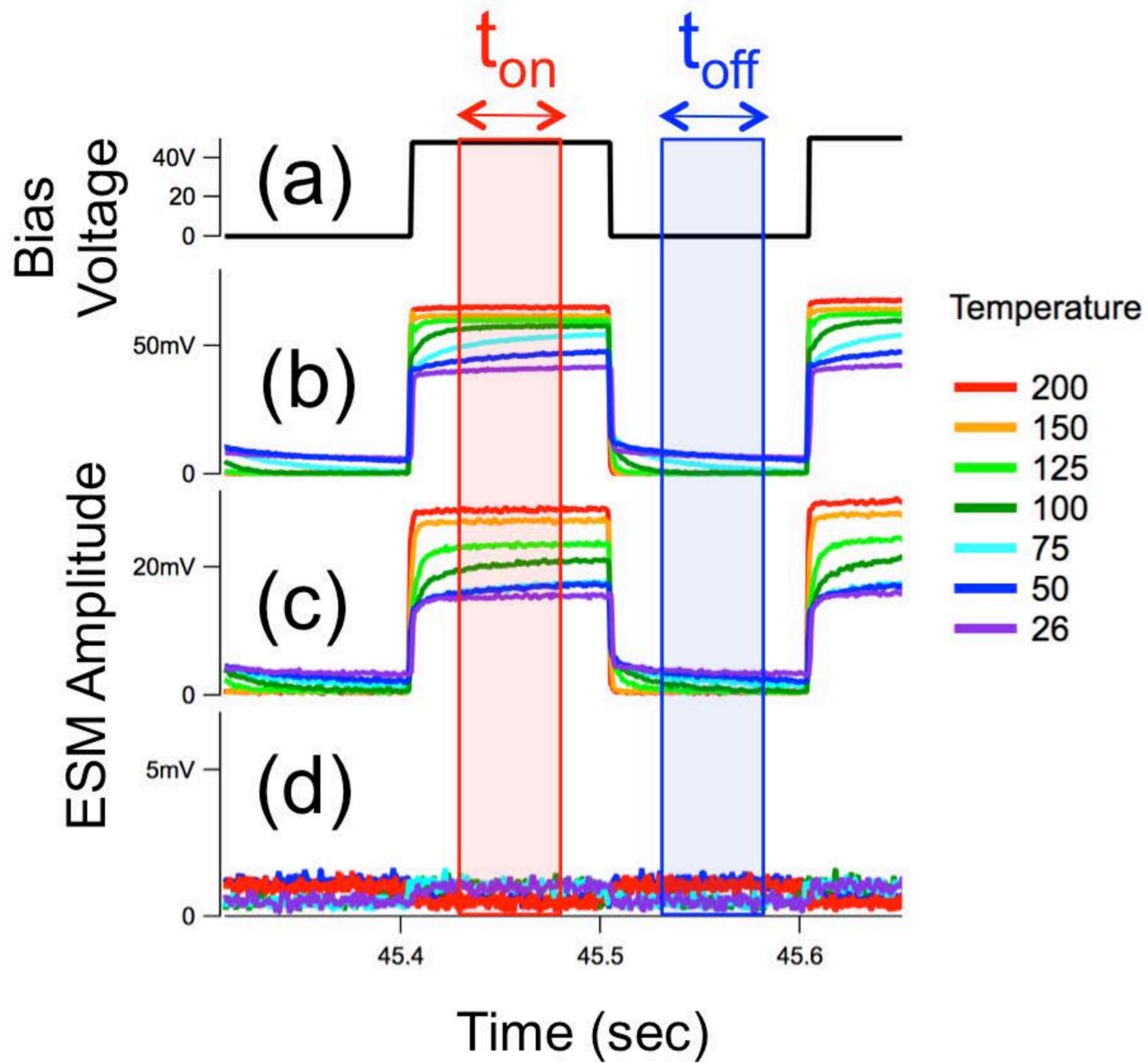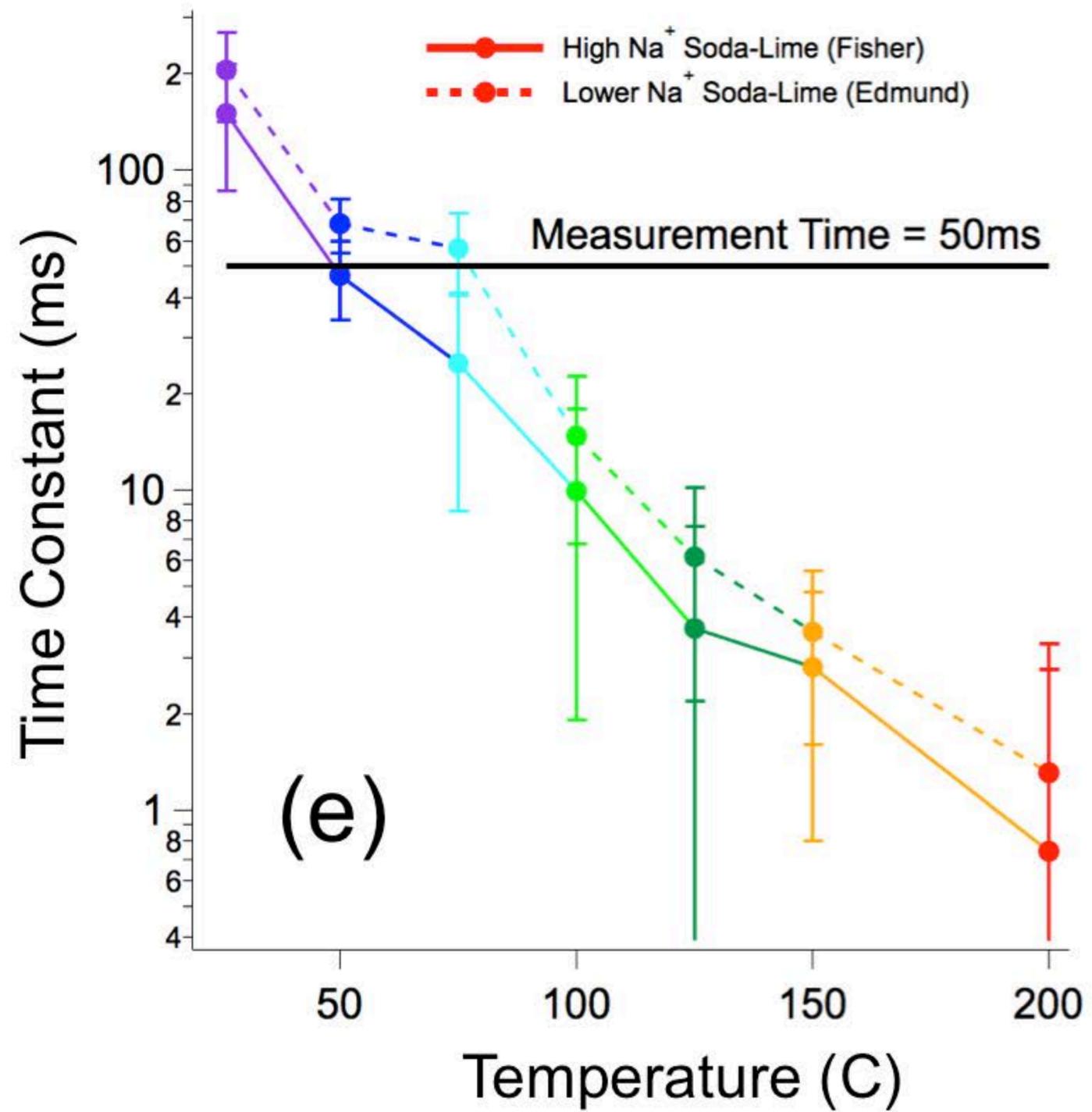

Figure 3, Proksch